\documentclass[aps,prb,twocolumn,superscriptaddress,floatfix,showpacs,amsmath,amssymb,nofootinbib,longbibliography,nobibnotes,noeprint]{revtex4-2}

\usepackage{graphicx}
\usepackage{dcolumn}
\usepackage{bm}
\usepackage{color}
\usepackage{tabularx}
\usepackage{csquotes}
\usepackage[normalem]{ulem}
\usepackage{xcolor}

\newcommand{\LiFeSeO}{(Li$_2$Fe)SeO}

\newcommand{\LiFeSO}{(Li$_2$Fe)SO}

\newcommand{\li}         {$^{7}$Li}

\newcommand{\lifeso}    {$({\mathrm{Li}}_{2} {\mathrm{Fe}}) {\mathrm{S}} {\mathrm{O}}$}
\newcommand{\lifeseo}   {$({\mathrm{Li}}_{2} {\mathrm{Fe}}) {\mathrm{Se}} {\mathrm{O}}$}

\newcommand{\slrrt}     {$(T_1T)^{-1}$}
\newcommand{\slrt}      {$T_1$}
\newcommand{\slrr}      {$T_1^{-1}$}

\newcommand{\ssrr}      {$T_2^{-1}$}
\newcommand{\degree}       {$^{\circ}$}

\newcommand{\bmed}{$B^\mathrm{median}_{\rm hyp}$}

\begin{document}


\title{Magnetic order and Li-diffusion in the
1/3-filled Kagome layers of antiperovskite Lithium-ion battery materials
(Li$_2$Fe)SO and (Li$_2$Fe)SeO}

\author{F. Seewald}\affiliation{Institut für Festk\"orperphysik, Technische Universit\"at Dresden, 01062 Dresden, Germany}

\author{T.~Schulze} \affiliation{Leibniz Institute for Solid State and Materials Research Dresden, 01069 Dresden, Germany} \affiliation{Technische Universit\"at Dresden, 01062 Dresden, Germany}

\author{N.~Gräßler} \affiliation{Leibniz Institute for Solid State and Materials Research Dresden, 01069 Dresden, Germany}

\author{F.~L. Carstens}
\affiliation{Kirchhoff Institute for Physics, Heidelberg University, INF 227, D-69120 Heidelberg, Germany}

\author{L.~Singer}
\affiliation{Kirchhoff Institute for Physics, Heidelberg University, INF 227, D-69120 Heidelberg, Germany}

\author{M.A.A. Mohamed} \affiliation{Leibniz Institute for Solid State and Materials Research Dresden, 01069 Dresden, Germany}
\affiliation{Department of Physics, Faculty of Science, Sohag University, 82524 Sohag, Egypt}

\author{S. Hampel} \affiliation{Leibniz Institute for Solid State and Materials Research Dresden, 01069 Dresden, Germany}

\author{B.~Büchner} \affiliation{Leibniz Institute for Solid State and Materials Research Dresden, 01069 Dresden, Germany} \affiliation{Technische Universit\"at Dresden, 01062 Dresden, Germany}

\author{R.~Klingeler}
\email{klingeler@kip.uni-heidelberg.de}
\affiliation{Kirchhoff Institute for Physics, Heidelberg University, INF 227, D-69120 Heidelberg, Germany}

\author{H.-H. Klauss}\email{henning.klauss@tu-dresden.de}\affiliation{Institut für Festk\"orperphysik, Technische Universit\"at Dresden, 01062 Dresden, Germany}

\author{H.-J.~Grafe}\email{H.Grafe@ifw-dresden.de}\affiliation{Leibniz Institute for Solid State and Materials Research Dresden, 01069 Dresden, Germany}
\date{\today}

\begin{abstract}
The recently discovered lithium-rich antiperovskites \LiFeSeO\ and \LiFeSO\ host lithium and iron ions on the same atomic position which octahedrally coordinates to central oxygens. In a cubic antiperovskite these sites form Kagome planes stacked along the $<111>$ directions which triangular motifs induce high geometric frustration in the diluted magnetic sublattice for antiferromagnetic interactions. Despite their compelling properties as high-capacity Li-ion battery cathode materials, very little is known about the electronic and magnetic properties of lithium-rich antiperovskites. We report static magnetization, Mössbauer, and NMR studies on both compounds. Our data reveal a Pauli paramagnetic-like behaviour, a long-range antiferromagnetically ordered ground state below $\approx$ 50~K and a regime of short-range magnetic correlations up to 100~K. Our results are consistent with a random Li-Fe distribution on the shared lattice position. In addition, Li-hopping is observed above 200~K with an activation energy of $E_{\rm a} = 0.47$~eV. Overall, our data elucidate static magnetism in a disordered magnetically frustrated and presumably semimetallic system with thermally induced ion diffusion dynamics.


\end{abstract}
\maketitle

\section{Introduction}

Lithium transition metal (TM) oxides are standard electrode materials for Li-ion batteries (LIBs)~(see, e.g.,~\cite{Manthiram2020,Du2022} and references therein). 
For their functionality the Li diffusion properties and Li intercalation capacity at room temperature and above are essential. Li-rich antiperovskite materials are promising candidates in this context (see, e.g.,~\cite{Zheng2021}). Recently, the class of (Li$_2$Fe)$Ch$O ($Ch =$~S, Se, Te) has been found~\cite{Lai.2017,Lu.2018,Deng.2021} and their feasibility as electrode materials in LIBs has become a current research topic in battery research. 

\begin{figure} [htb] 
    \includegraphics[width = 1\columnwidth]{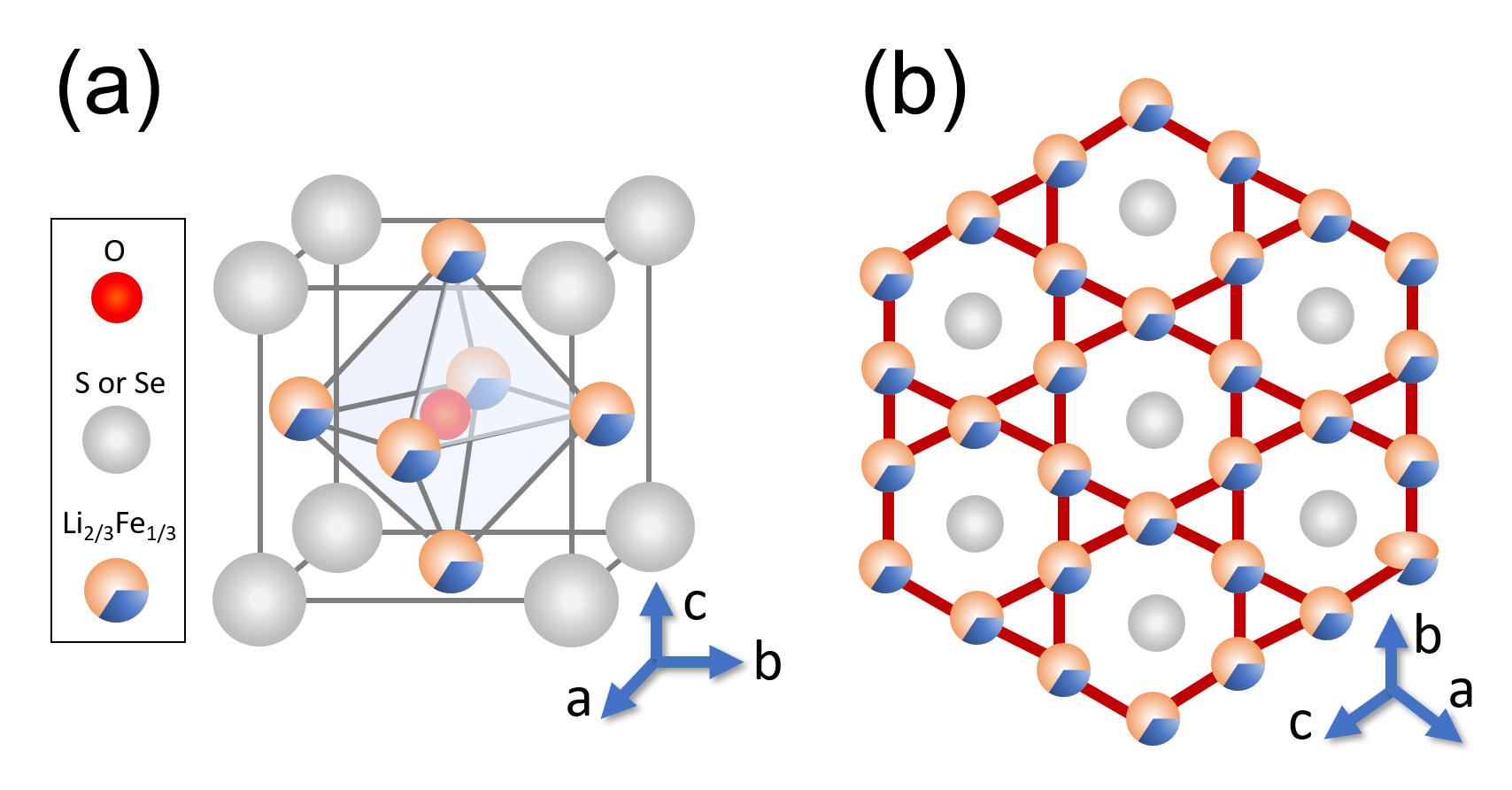}
    \caption{(a) Schematics of the crystallographic unit cell of (Li$_2$Fe)ChO (Ch = Se or S) (after~\cite{Lai.2017}). 1/3 (2/3) of the crystallographic 3$c$ sites of the $Pm\bar{3}m$ structure are filled with paramagnetic Fe (diamagnetic Li) species. (b) View along the $<111>$   direction; red lines between the 3$c$ sites illustrate the underlying Kagome layers.
    }
    \label{unitcell}
\end{figure}

Li-rich antiperovskites posses cubic crystal structures (space group $Pm\bar{3}m$~\cite{Lai.2017}) as sketched in Fig.~\ref{unitcell}(a), in which lithium and iron ions occupy the same lattice position (3c) with 2/3 and 1/3 probability, respectively. Since Li$^+$ and Fe$^{2+}$ are too small for their common atomic position they exhibit large thermal displacements implying a high cation mobility~\cite{Lai.2017} which is a prerequisite for their use in LIBs. In theoretical studies of \LiFeSO\ a thermally activated Li-diffusion with a very small diffusion barrier of 0.32~eV was deduced \cite{Lu.2018}. For ionic diffusion, disorder can have dramatic effects. In  most Li-TM oxides disorder is caused by various defects which indeed have an enormous effect on Li diffusion by blocking or enabling low-energy ionic pathways~\cite{Wang2007,Malik2010,Reynaud2023}
and on magnetism~\cite{Grafe2013,Werner2020,Ranaut2023,Goon2023,Zeng2025}. In contrast, randomness is much more pronounced in Li-rich antiperovskites due to the shared Li-TM position. It produces long-range disorder, yet a preferential oxygen coordination has been recently suggested to stabilize polar short-range cation orderings~\cite{Coles2023,Deng2023}. 

From the viewpoint of fundamental research in magnetism, the presumably largely random distribution of diamagnetic and paramagnetic species on the same lattice site render Li-rich antiperovskites model systems to study magnetism and the potential evolution of long-range magnetic order in a disordered semimetal. The TM sites in antiperovskites exhibit Kagome planes stacked along the $<111>$ directions (see Fig.~\ref{unitcell}(b))~\cite{Singh21}. Therefore, for nearest neighbor antiferromagnetic interaction strong magnetic frustration on triangular units is expected. Since the Kagome planes exist along all $<111>$ directions this is not a two-dimensional but an isotropic three-dimensional magnetic system. No macroscopic magnetic ground state degeneracy is expected. 

In this work we identify and characterize \LiFeSeO\ and \LiFeSO\ with respect to magnetic order of the Fe moments and examine if the Li and Fe ions are indeed statistically distributed on the shared lattice site. Moreover we search for signatures of Li diffusion in \li\ NMR spectroscopy and evaluate the experimental energy barrier for Li diffusion which is determined as $E_{\rm a}\simeq 0.47$~eV. Notably, long range antiferromagnetic order is observed in our macroscopic and local probe studies which in particular enable us to determine the temperature dependence of the magnetic order parameter. While long-range antiferromagnetic order appears below 50~K, short-range order is observed up to $\sim 100$~K. This reflects the strong dilution of the iron ions on the TM site as well as the partly lifted geometric frustration of the antiferromagnetic nearest neighbor magnetic exchange interaction within the Kagome planes. The geometric frustration manifests itself also in a non-linear dependence of the $^{57}$Fe magnetic hyperfine field on the number n of nearest neighbor Fe ions in the long-range ordered state.

\section{Experimental Methods}

The \LiFeSO\ and \LiFeSeO\ samples have been synthesized by ball milling (BM)~\cite{Mohamed2023}. In addition to investigating the pristine materials, post-synthesis heat-treatment at 300 \degree C and 500 \degree C, respectively, has been applied to further reduce the content of impurity phases~\cite{Singer2024}. Details of the synthesis as well as a detailed characterization of the \LiFeSO\ and \LiFeSeO\ samples can be found in~\cite{Singer2023, Mohamed2023}. 

Magnetic measurements were performed on powder samples using a MPMS3 magnetometer (Quantum Design). The static magnetic susceptibility $\chi =M/B$ has been obtained upon varying the temperature at 1~T by using field-cooled (FC) and zero-field-cooled (ZFC) protocols where the sample was cooled either in the external measurement field or the field was applied after cooling to the lowest temperature. Isothermal magnetisation $M(B)$ has been measured at $T=1.8$~K in the field range -7~T~$\leq B\leq$~+7~T.

$^{57}$Fe-Mössbauer measurements have been conducted in a CryoVac flow cryostat under He atmosphere.
A Rh/Co source driven by a Mössbauer WissEL drive unit MR-360 biased by a DFG-500 frequency generator in sinusoidal mode was used.
The detection device is a proportional counter tube in combination with a CMTE multichannel data processor MCD 301/8K and a WissEL single channel analyzer Timing SCA to set the energy window.
Powder samples of \LiFeSO \
and \LiFeSeO \ 
were investigated.
All data evaluation has been performed using Moessfit \cite{Moessfit}, using the implemented maximum entropy method (MEM) to describe fit parameter distributions.
Measurements have been carried out between 4.2\,K and 300\,K .
All center shifts are stated relative to room temperature $\alpha$-Fe.
The linewidth $\omega$ is given as half width at half maximum (HWHM).

\li\ NMR has been measured in an external magnetic field of 2.001~$T$. Due to a gyromagnetic ratio of $\gamma = 16.5471$~MHz/T  for the \li\ isotope the resonance frequency for the bare nucleus is 33.11~MHz. The \li\ spectra have been taken by the Hahn-Spin-Echo sequence. Since the spectra broaden at low temperatures, we applied a step-and-sum technique, i.e. we measured spectra while stepping the frequency in 50~kHz steps and added the Fourier-transformed spectra taken at each step. The spin-lattice relaxation time, \slrt , has been measured at the peak position of the spectra by a saturation recovery sequence using a train of 90\degree\ pulses at a time $t$ before a Hahn-Spin-Echo sequence. The spin-spin relaxation time, $T_2$, has been measured at the peak of the spectra by varying the time $\tau$ between the 90\degree\ and the 180\degree\ pulse of the Hahn-Spin-Echo sequence.

\section{Results and Discussion}

\subsection{Magnetic Susceptibility and Magnetization}
\label{sec:Magprop}

\begin{figure} [htb] 
    \includegraphics[width = \columnwidth]{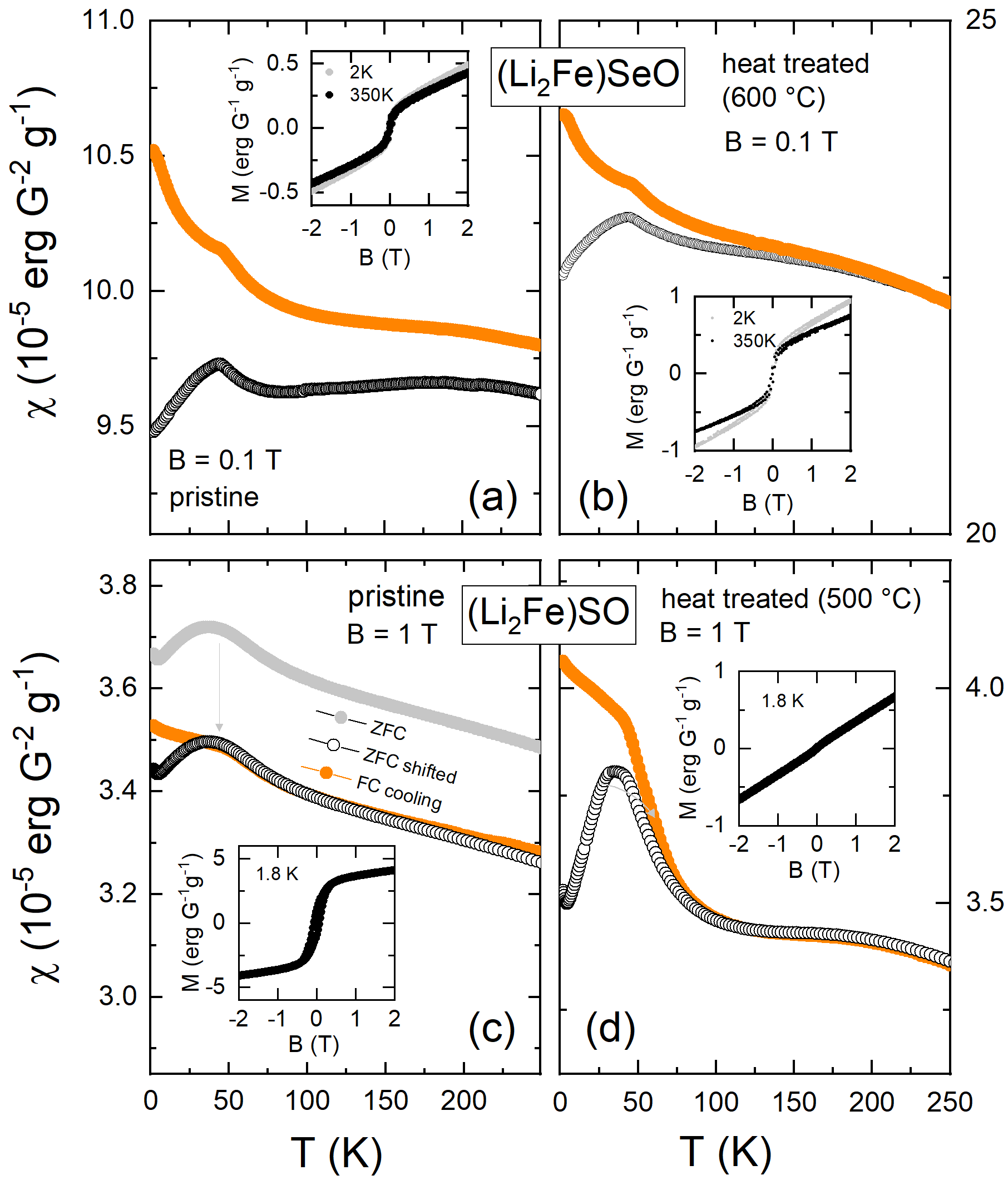}
    \caption{Static magnetic susceptibility ($\chi = M/B$) of (a,b) \LiFeSO\ and (c,d) \LiFeSeO\ vs. temperature up to 250~K. Both FC (filled orange markers) and ZFC (open black markers) protocols have been applied. (a,c) show pristine materials, (b,d) post-synthesis heat-treated materials. In (c), ZFC data (grey) have been shifted (black) to compensate for a hysteresis effect of a small ferromagnetic impurity phase. Insets: Corresponding isothermal magnetization. 
    }\label{SQUIDall}
\end{figure}

In Fig.~\ref{SQUIDall} we depict the static magnetic susceptibility $\chi(T) = M(T)/B$ using zero field cooled (ZFC) and field cooled (FC) protocols as well as isothermal magnetization measurements $M(B)$ of two samples of \LiFeSeO\ (as-grown and with post-synthesis 600 \degree C heat treatment, respectively) and two samples of \LiFeSO\ (as grown and with 500 \degree C  heat treatment). In the ZFC data, all samples exhibit a maximum in $\chi(T)$ as well as a FC/ZFC splitting.  Upon heating between 100\,K and 300\,K we find a nearly constant or weakly decreasing (typically by 5~\%) magnetic susceptibility which is typically associated with strong Pauli paramagnetism of 3$d$ conduction electrons. Below 100\,K all samples exhibit an increase of $\chi (T)$ which -- by comparison with the local probe data presented below -- we attribute to the onset of short range magnetic correlations. The as-grown samples of both systems show anomalies which are associated with ferromagnetic minority phases with typical 3~\% volume fraction. We suspect this to be iron and Fe$_x$S with $x \leq 1$ which has been detected in x-ray diffraction and Moessbauer spectroscopy. The presence of ferromagnetic impurity phases yields a small offset in the $\chi$ vs. $T$ curves as well as corresponding magnetization steps at zero external field in the magnetization measurements which are absent, e.g., in the heat-treated \lifeso\ (see insets of Fig.~\ref{SQUIDall}). A nearly temperature-independent magnetic susceptibility, however at 5 times larger absolute values of $\chi$ and without clear low-temperature features as observed here, has been reported for a \lifeso\ sample with larger ferromagnetic, possibly Fe$_3$O$_4$, impurity contents~\cite{Mikhailova2018,Singer2024}. The effect of heat-treatment on reducing and eliminating such magnetic impurity phases in \lifeso\ is discussed in Refs.~\cite{Singer2024,Mohamed2023}.

\begin{figure} [htb] 
    \includegraphics[width = \columnwidth]{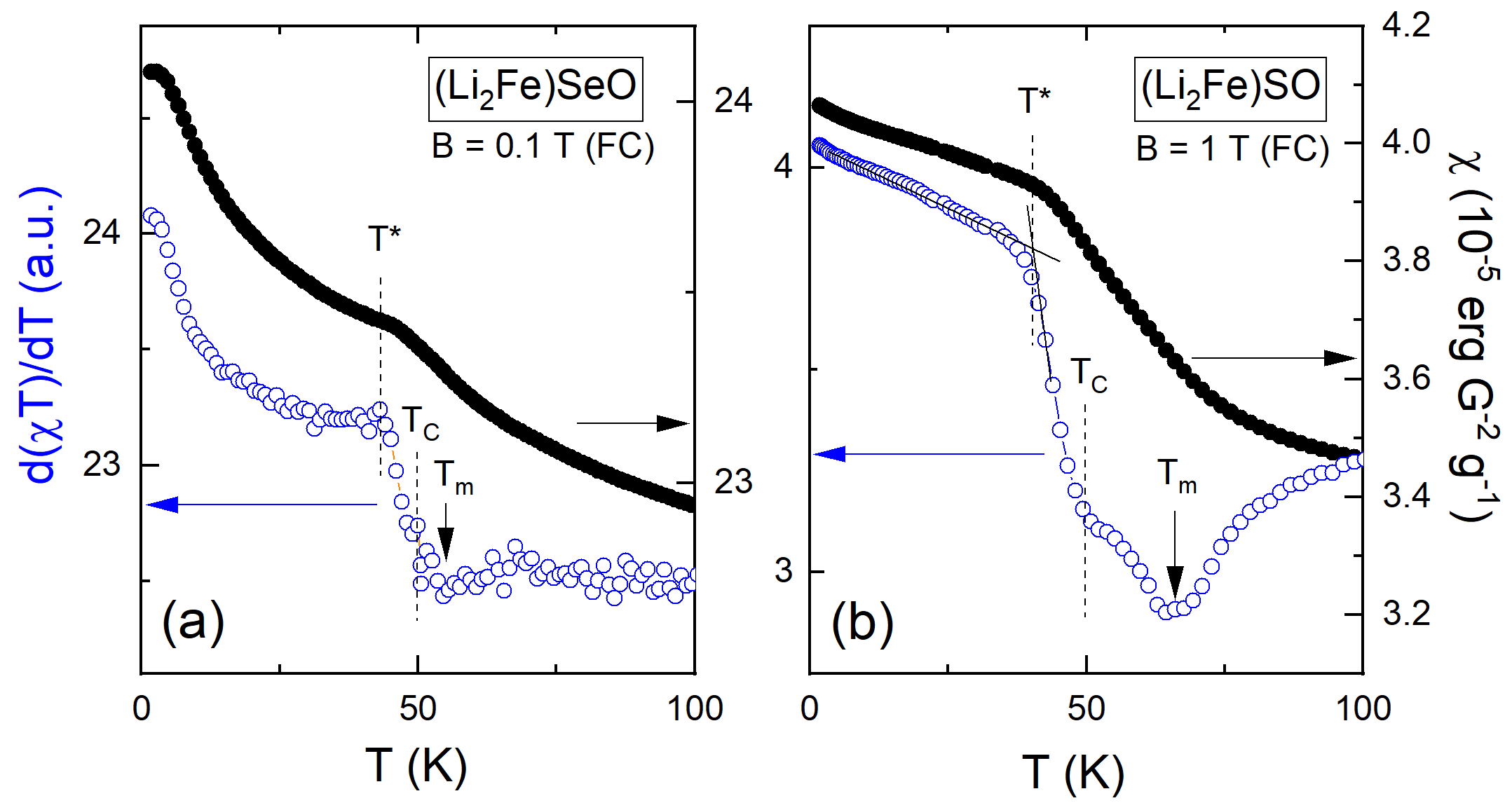}
    \caption{Fisher's specific heat $\partial (\chi T)/\partial T$ (blue open markers) as derived from the field cooled static magnetic susceptibility (black filled markers) in Fig.~\ref{SQUIDall}b,d. The shoulder/minimum feature marked by $T_{\rm C}$ signals the evolution of long-range magnetic order as evidenced by the local probe data (see Fig.~\ref{MBS-FIG:BHypS}). $T^*$ marks a kink and $T_{\rm m}$ a further minimum in Fisher's specific heat (see the text). 
    }
    \label{cpmag}
\end{figure}

The FC measurements of the heat-treated materials allow us to further investigate the evolution of magnetic order by considering Fisher's specific heat $\partial (\chi T)/\partial T$ which is derived from the static magnetic susceptibility data. This quantity is proportional to the magnetic specific heat~\cite{Fisher1962}. In both materials, Fisher's specific heat indicates a broad regime of magnetic entropy changes extending from $\simeq 30$~K to above 100~K (see Fig.~\ref{cpmag}). In \lifeseo\ as well as in \lifeso , a step-like feature appears at around 50~K which indicates a magnetic phase transition. As will be shown by our local probe data below (e.g. documented by the appearance of static magnetic hyperfine fields at the $^{57}$Fe nuclei below 60~K in Fig. \ref{MBS-FIG:4-295K-S-Se}), this feature is associated with a long-range antiferromagnetic order transition of the Fe moments. Towards lower temperatures, the step in the magnetic specific heat is contained by a tiny maximum respectively a kink at $\simeq 44$~K marked by $T^*$ in Fig.~\ref{cpmag}. We also note the existence of a minimum in $\partial (\chi T)/\partial T$ for \lifeso\ at 65~K which indicates a precursing magnetically ordered phase~\cite{Spachmann2022l}.

\subsection{Moessbauer spectroscopy}

\begin{figure} [htb]
		\begin{center}
		\includegraphics[width=\columnwidth]{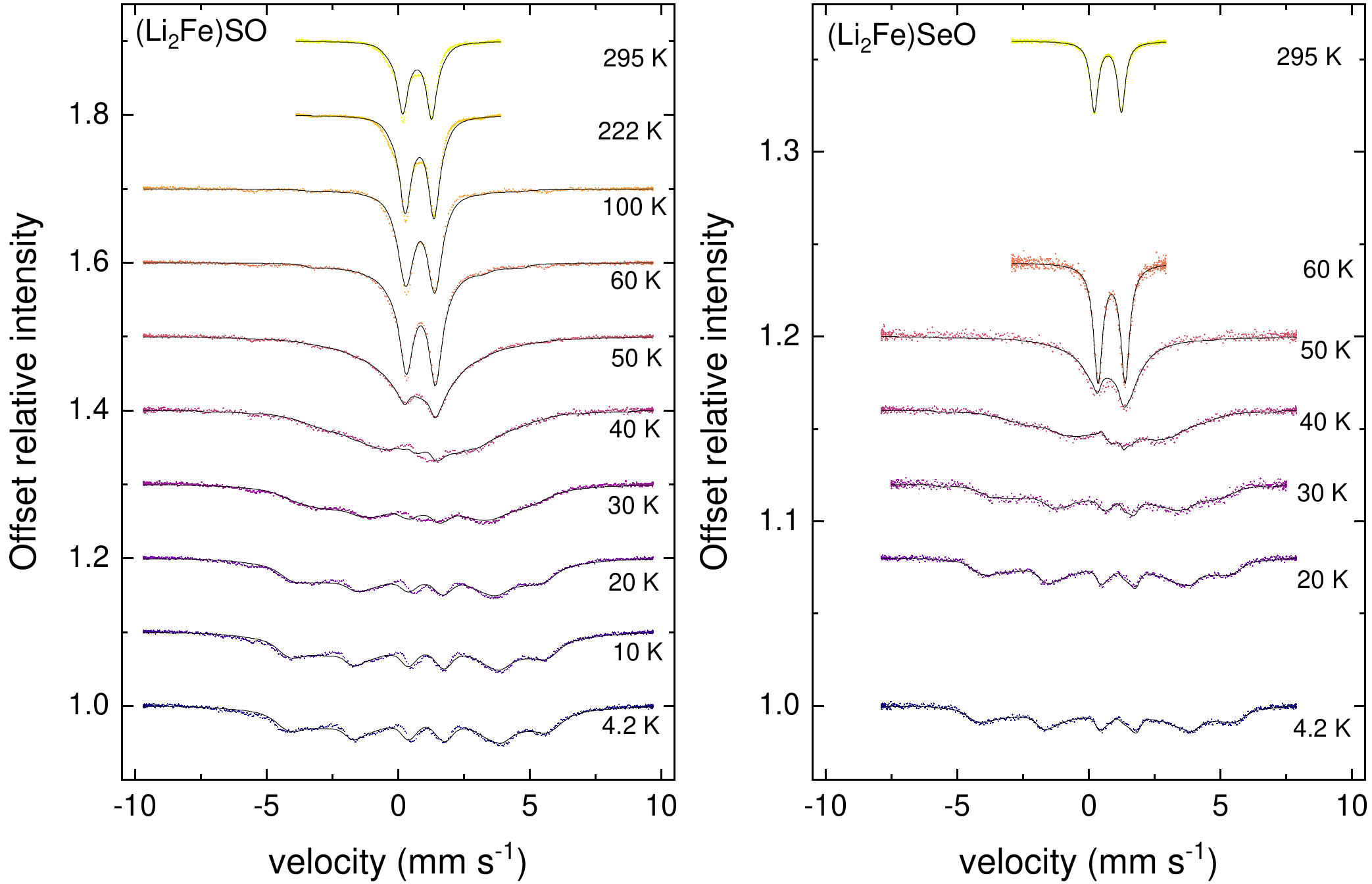}
		\caption{Mössbauer spectra of \LiFeSO \ and \LiFeSeO \ between 4.2\,K and 295\,K. Below 60~K the onset of static magnetic order is documented by a spectral broadening and the gradual appearance of a six line magnetic hyperfine field splitting.}
		\label{MBS-FIG:4-295K-S-Se}
        \end{center}
	\end{figure}

\begin{figure} [htb]
\centering
\includegraphics[width=1\columnwidth]{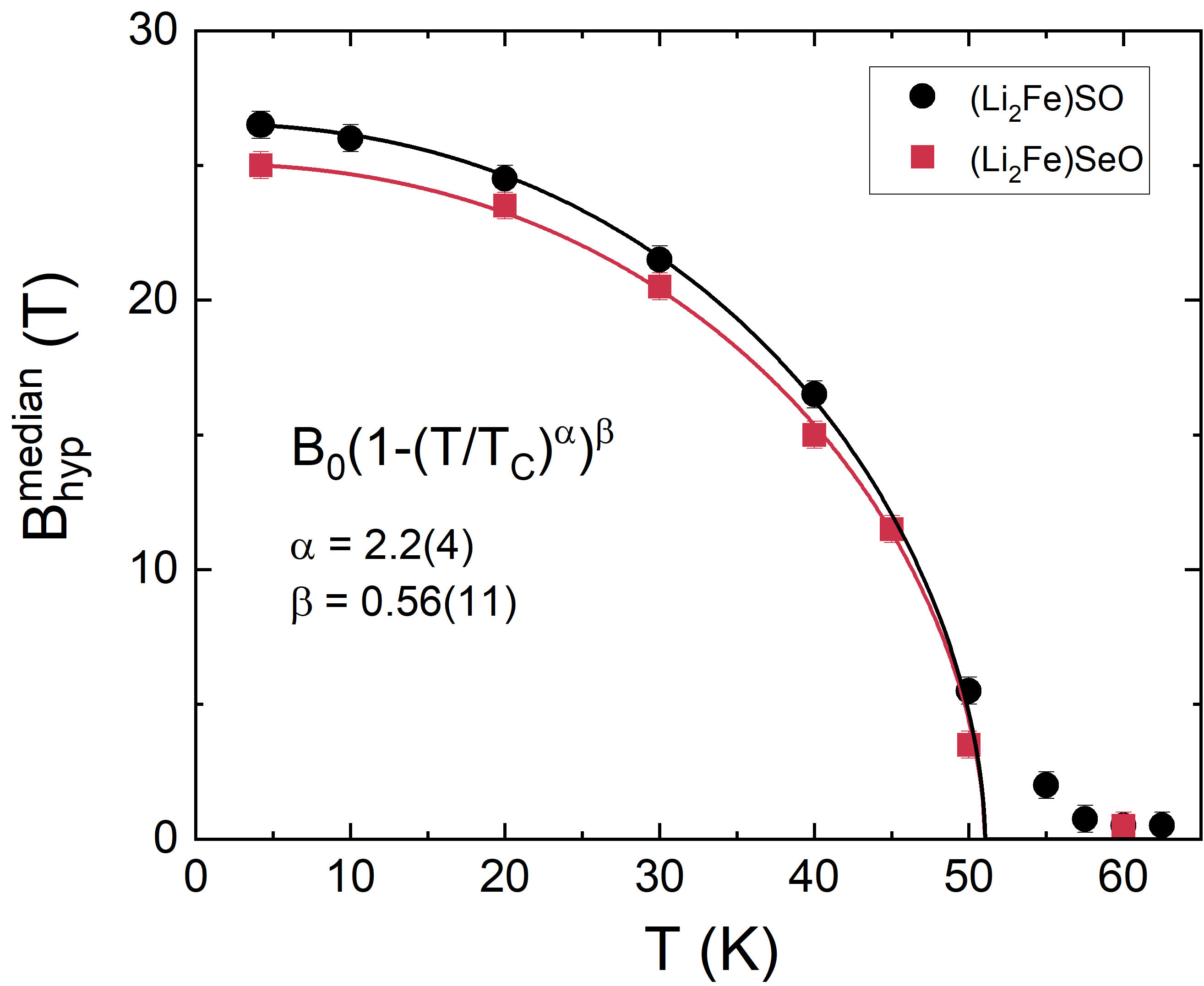}
\caption{Temperature dependence of the median magnetic hyperfine field \bmed\  
for \LiFeSO \ (black squares) and \LiFeSeO \, (red dots).
The solid lines describe a phenomenological model \bmed~$= B_\mathrm{0}\times (1-(\frac{T}{T_\mathrm{C}})^\alpha)^\beta$ with shared $T_\mathrm{C}$, $\alpha$ and $\beta$  (see the text).}
\label{MBS-FIG:BHypS}
\end{figure}

Figure\,\ref{MBS-FIG:4-295K-S-Se} depicts selected $^{57}$Fe-Mössbauer powder spectra of \LiFeSO\ and \LiFeSeO.
At room temperature the spectra of both samples can be described with a single main iron site exhibiting quadrupole splitting due to a finite electric field gradient (EFG) at the $^{57}$Fe nucleus. Below 60~K the spectra broaden and six absorption lines develop and document the gradual appearance of an additional static magnetic hyperfine field at the $^{57}$Fe nucleus. As will be discussed in detail below the magnetic hyperfine field distribution is not homogeneous and single-valued yet broadened due to the probability distribution of the number of Fe-Fe neighbors on the Fe site. Except a minor precursor observed for  \lifeso\ at 55~K the onset and continuous increase of static magnetic hyperfine fields for the full spectrum is observed in both systems at 50~K and below. This order parameter-like increase (see Fig.~\ref{MBS-FIG:BHypS}) proves a magnetic phase transition into a magnetically long-range ordered state. 

\begin{figure} [htb]
		\begin{center}
		\includegraphics[width=\columnwidth]{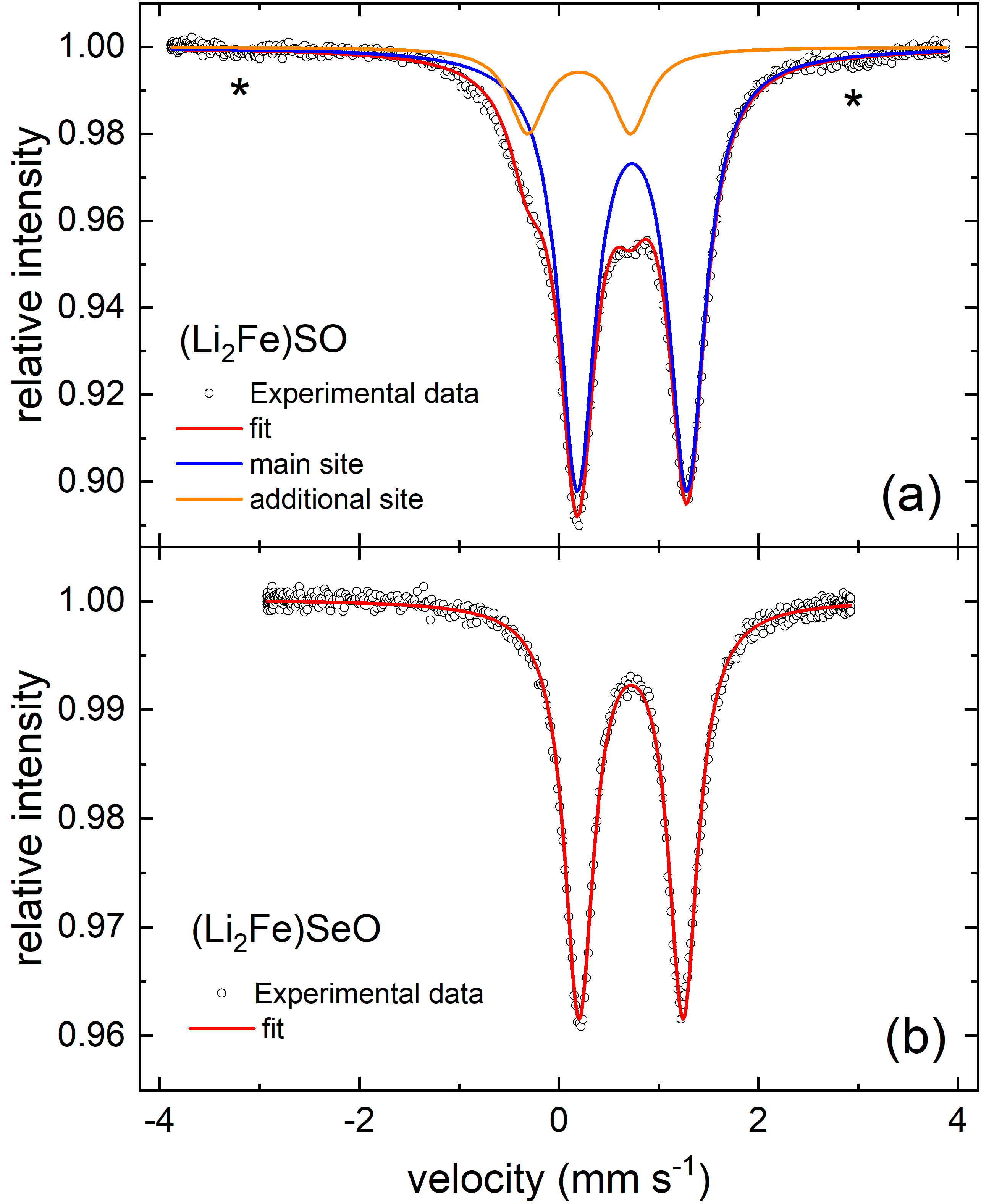}
		\caption{Mössbauer spectrum of \LiFeSO \ (a) and \LiFeSeO \ (b) at room temperature. For \LiFeSO , in addition to the main site a second quadrupole split doublet with hyperfine parameters similar to Fe$_x$ and $\alpha$-LiFeO$_2$ is modeled to account for the main doublet asymmetry. Moreover, two weak magnetic absorption lines are observed at -3.2~mm\,s$^{-1}$ and +2.8~mm\,s$^{-1}$ marked by asterisks which are associated with a $<5$~\% $\alpha$-iron impurity. The \LiFeSeO \ spectrum is described by a single quadrupole split site.}
		\label{MBS-FIG:RoomTempS-Se}
        \end{center}
	\end{figure}

At room temperature the spectra of both samples can be described with a single main iron site exhibiting quadrupole splitting as seen in Fig.\,\ref{MBS-FIG:RoomTempS-Se}.
Assuming an EFG with principle components $\|V_{zz}\| \geqq \| V_{yy} \|\geqq \| V_{xx} \|$ and an asymmetry of $\eta = \frac{V_{xx}-V_{yy}}{V_{zz}} = 0$ the value obtained for the main principle component $V_{zz}$ is $V_{zz}$ = 65.72(20)~V\,Å$^{-2}$ and $V_{zz}$ = 62.7(12)~V\,Å$^{-2}$ for the sulfur and selenium samples respectively.
In \LiFeSO\ a temperature dependent asymmetry between the two peaks as well as pronounced shoulders are observed in addition to the main powder site. 
To describe the significant asymmetry of the doublet observed for the sulfur sample an additional doublet with 16.4(14)~\% spectral area, an EFG principle component of $V_{zz}$  = 62.2(25)~V\,Å$^{-2}$, and a center-shift of 0.313(20)~mm\,s$^{-1}$ is used. This can be associated with Fe$_x$S with $x \leq 1$ or LiFeO$_2$ which have been detected as minority phases in XRD.
The shoulders at $\approx$ -3~mm\,s$^{-1}$ and +2.5~mm\,s$^{-1}$ could be associated with a small volume (less than 5~\%) FeS magnetic impurity phase. The linewidth of 0.218(16)~mm\,s$^{-1}$ for \LiFeSO\ and 0.184(16)~mm\,s$^{-1}$ for  \LiFeSeO\ prove that there is only a small variation of $V_{zz}$ for different local iron environments resulting from the iron lithium disorder.

\begin{figure} [htb]
        \begin{center}
        \includegraphics[width=\columnwidth]{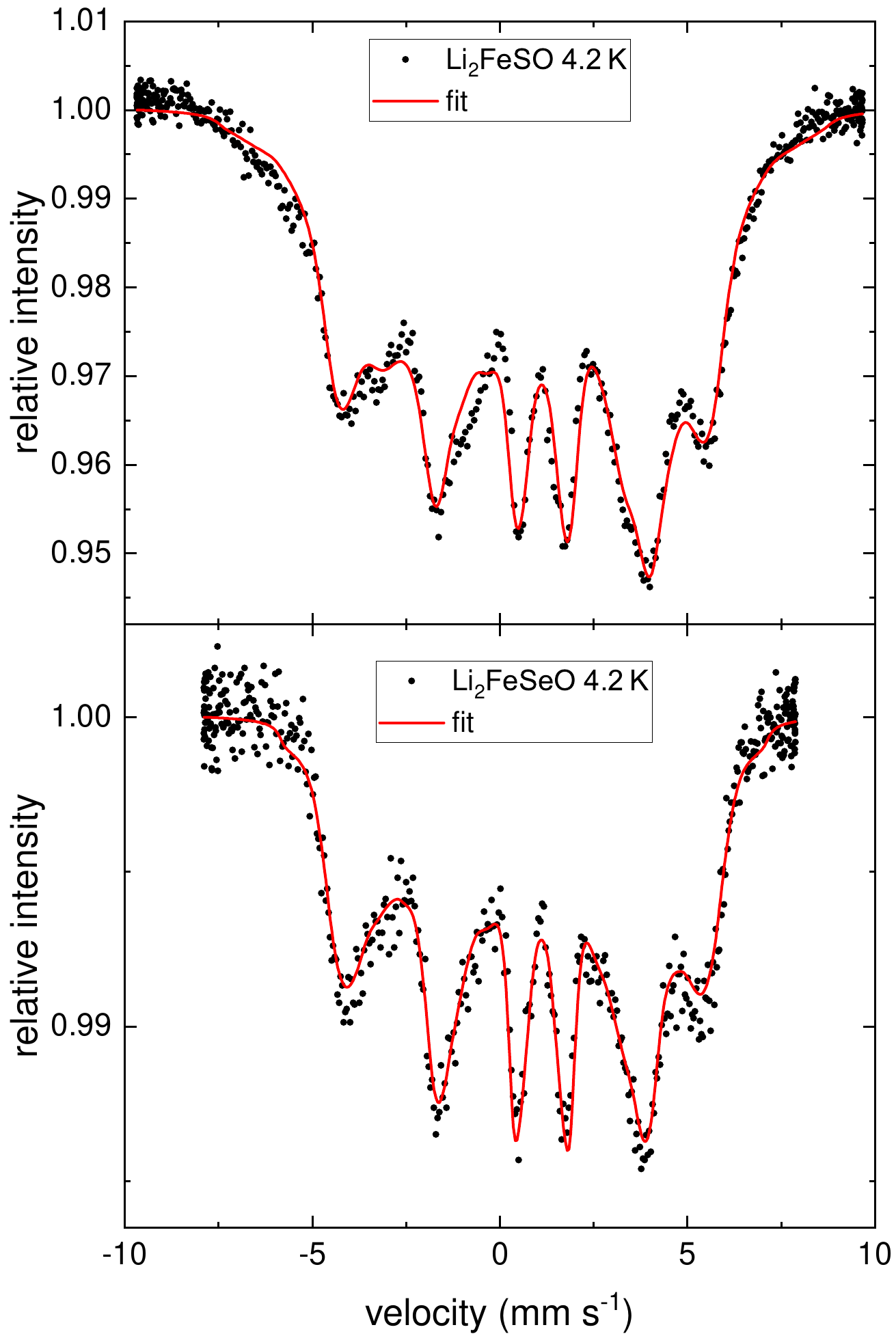}
        \caption{Mössbauer spectrum of \LiFeSO\ (upper panel) and \LiFeSeO\ (lower panel) at 4.2~K. A magnetic sextet with broad lines is observed and modeled by a hyperfine field distribution.}
        \label{MBS-FIG:4.2SandSe}
        \end{center}
	\end{figure}

\begin{figure} [htb]
		\begin{center}
		
        \includegraphics[width=\columnwidth]{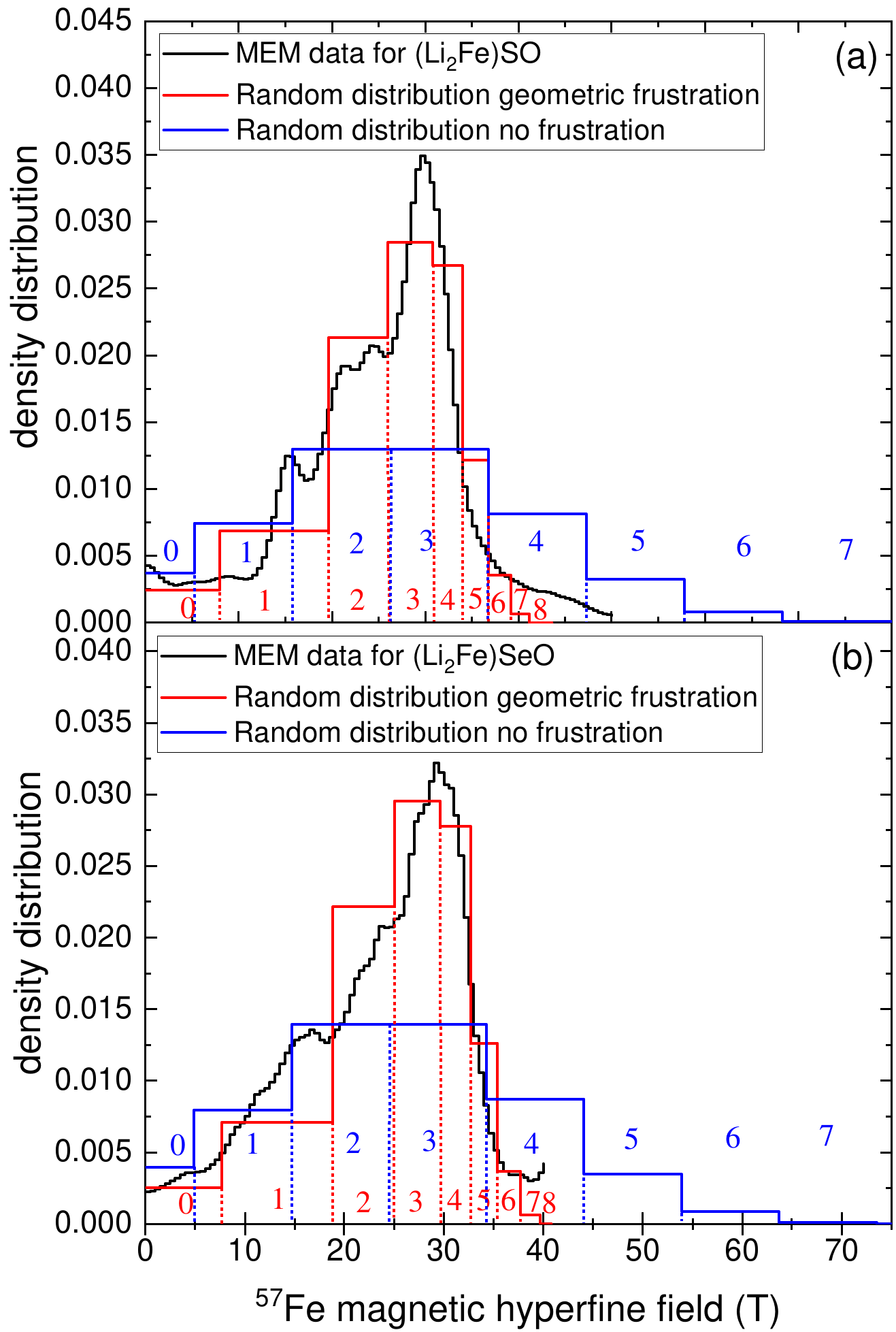}
		\caption{(black lines) Measured magnetic hyperfine field probability distribution $\rho$ in \LiFeSO\  (a) and  \LiFeSeO\ (b)  at 4.2~K  compared with two different models assuming a random distribution of iron and lithium ions on the shared lattice  site: (blue lines) linear model (unfrustrated exchange), (red lines) frustrated exchange model with B$(i) \propto 1/i$. All probability distributions are normalized to a total area of 1. 
        Blue and red numbers indicate the number $i$ of nearest neighbor iron ions for the two models. 
        }
		\label{MBS-FIG:parameterdist4.2KS-Se}
        \end{center}
	\end{figure}

At 4.2~K both samples show inhomogeneously broadened sextets proving the presence of local magnetic hyperfine fields $B_\mathrm{hyp}$ at the iron positions (see Fig. \ref{MBS-FIG:4.2SandSe}). The strongly increased linewidth with respect to the non-magnetic high temperature spectra can be attributed to a distribution of hyperfine fields at different iron atoms. Fig.\,\ref{MBS-FIG:parameterdist4.2KS-Se} presents the measured distribution of local hyperfine fields at 4.2~K, with the median values of 26.25~T for the sulfur sample and 24.75~T for the selenium sample. The spectra are deduced from a maximum entropy method (MEM) analysis with a $B_{\mathrm{hyp}}$ resolution of 0.5~T. The observation of a hyperfine field distribution is consistent with  Li-Fe disorder on the shared lattice position since this leads to a variety of local iron environments with varying numbers of nearest iron neighbors.
    
Figure~\ref{MBS-FIG:parameterdist4.2KS-Se} also depicts two calculated hyperfine field distributions. Recent X-ray pair distribution function (PDF) studies have shown a nearly random occupation of the Li/Fe shared lattice position~\cite{Coles2023,Deng2023}. Each iron has 8 nearest TM neighbors with 2/3 lithium and 1/3 iron occupancy. Table \ref{Tab-MBS:randomFe-Li distribution} displays the relative probabilities $f_n$ for $n$ Fe-O-Fe neighborhoods assuming a random lithium/iron distribution given by 
    \begin{equation}
            f_n = \frac{8!}{(8-n)!n!} \times (\frac{1}{3})^n \times (\frac{2}{3})^{8-n}.
            \label{eq:randLiFe}
    \end{equation}
    \begin{table}[h] 
		\begin{center}
		\begin{tabular}{c|c|c|c|c|c|c|c|c|c}
			No. of Fe neighbors  & 0 & 1 & 2 & 3 & 4 & 5 & 6 & 7 & 8\\\hline
			Probability [\%] & 3.9 & 15.6 & 27.3 & 27.3 & 17.1 & 6.8 & 1.7 & 0.2 & 0.0
		\end{tabular}
		\caption{Probability for the various possible numbers of nearest iron neighbors assuming a random Fe-Li distribution.}
		\label{Tab-MBS:randomFe-Li distribution}
        \end{center}
	\end{table} 
 The Fe magnetic hyperfine field is proportional to the local magnetic exchange field at the Fe ion. To model the distribution we consider the number of Fe nearest and next-nearest neighborships in the lattice. Due to the disordered nature of the magnetic lattice we assume that local superexchange interactions dominate.  The 8 nearest neighborships involve 
  a 60\degree\ antiferromagnetic exchange via two (S,Se) ions and a weak ferromagnetic 90\degree \, exchange via an oxygen ion. 
 Due to the antiferromagnetic character of the long-range magnetic transition at $\approx$ 50~K we consider an effective antiferromagnetic exchange with the 8 nearest neighbor Fe ions. 
 In an unfrustrated magnetic lattice the resulting local hyperfine field  scales with the number $n$ of Fe neighbors, i.e. $ B(n) = B_0 \times n $ with $B_0$ being a scaling parameter of the hyperfine field chosen to match the experimental maximum. $B_0$ and the corresponding fraction $f_n$ obtained from Eq.~\ref{eq:randLiFe} define the area of the corresponding rectangle in Fig.~\ref{MBS-FIG:parameterdist4.2KS-Se}. The resulting hyperfine field distribution is plotted in blue color.  This distribution shows a broad high field shoulder not seen in the experimental spectrum. A possible interpretation would be that Fe-Fe configurations with 5, 6, or 7 nearest neighbors are not realized and the random Li-Fe distribution assumption (in which approx. 9~\% Fe sites are expected for these configurations) is not realized. However, this would contradict the interpretation of recent X-ray PDF studies \cite{Coles2023,Deng2023}. 
 
 Considering that the dominant local magnetic exchange geometry is triangular with strong geometric frustration for antiferromagnetic exchange, the local magnetic exchange field observed via the Fe magnetic hyperfine field is not linear in the number $n$ of nearest neighbors but exhibits a saturation for increasing $n$. Therefore we model $B(n)$ using the phenomenological ansatz
 \begin{equation}
 B(n) = B_0 \times \sum_{i=1}^n \frac{1}{i},
 \end{equation}
in which the contribution of each additional nearest neighbor $i$ will increase the total exchange field by a contribution proportional to $1/i$. The resulting hyperfine field distribution (red color in Fig.~\ref{MBS-FIG:parameterdist4.2KS-Se}) describes the asymmetric experimental spectrum  in \LiFeSO\   and  \LiFeSeO\  at 4.2~K much better than the linear model.

Note, in this figure we also include a contribution from the 3.9~\% Fe sites which do have no nearest Fe neighbor but may experience an exchange field due to second nearest Fe neighbors. The presence of such higher order interaction terms is indeed suggested by the fact that our $M(B)$ data of thermally treated \lifeso\ implies only an insignificant amount of quasi-free moments, i.e., obeying a Brillouin-like behavior, in the 1\textperthousand-regime (see Fig.~\ref{SQUIDall}d). For the Fe sites without nearest paramagnetic neighbors we assume a flat field distribution in the field range between 0 T and the onset field value of the model distribution for finite numbers of nearest Fe neighbors.

The MEM fit model can be employed to describe the spectra at all temperatures. Fig.\,\ref{MBS-FIG:BHypS} shows the temperature dependent median magnetic hyperfine fields \bmed\ 
following a $B_\mathrm{hyp} = B_0 \times (1-(\frac{T}{T_c})^\alpha)^\beta$ behavior with $\alpha$ = 2.2(4) and $\beta$ =0.56(11) \cite{PhysRevB.66.144429}.
This phenomenological model is adapted to the data assuming identical $T_\mathrm{C}$, $\alpha$ and $\beta$  for both samples scaled by independent low temperature saturation values $B_\mathrm{0}$ = 26.54(38)~T (\LiFeSO) and $B_\mathrm{0}$ = 25.1(5)~T (\LiFeSeO).
In this model $\beta$ is the usual critical exponent of the order parameter in the Wilson renormalization group theory describing the decrease close to the critical temperature and $\alpha$ describes the change of the order parameter close to $T = 0$~K. 
Hence, the deduced value for  $\beta$ is consistent with 0.5 which is the mean field value, and $\alpha$ is slightly smaller than 3 which is the exponent for the Bloch long wavelength magnon-induced order parameter decrease for antiferromagnets close to T=0.
    
For the selenium sample 
\bmed\ tends towards 0 above 50~K and spectra recorded at $T = $ 55~K and above are compatible with setting the maximum hyperfine field to 0.
For the sulfur sample \bmed\ 
tends to 0 between 55~K and 60~K. This observation corresponds to the minimum in Fisher's specific heat which indicates an anomaly in the magnetic entropy changes at 65~K (see Fig.~\ref{cpmag}). Such a feature above $T_{\rm C}$ is absent in the Fisher specific heat of \lifeseo .
As depicted in fig. \ref{MBS-FIG:RoomTempS-Se} (upper panel) a small fraction of iron nuclei (less than 5~\%) experiences a nonzero local magnetic hyperfine field up to room temperature. This can be associated with an $\alpha$-iron impurity phase.
	
	Introducing the angle $\theta$ between the local hyperfine field and the principle EFG component as a free parameter enables us to describe the data with global values for $\theta$ and $V_{zz}$ shared for all temperatures.
    For the sulfur sample we obtain $V_{zz, \mathrm{global}} = $ 65.13~V\,Å$^{-2}$ and $\theta_{\mathrm{global}} = 68.59^\circ$ and $V_{zz, \mathrm{global}} = $ 61.97~V\,Å$^{-2}$ and $\theta_{\mathrm{global}} = 79.73^\circ$  for the selenium sample.
	In this model the angle $\theta_{\mathrm{global}}$ is an effective angle averaged over different local iron environments due to the iron-lithium disorder.

\subsection{NMR results}
\label{sec:NMR}

\subsubsection{Analysis of NMR spectra}
\label{sec:NMRspectra}

\begin{figure}
\begin{center}
\includegraphics[width=\columnwidth,clip]{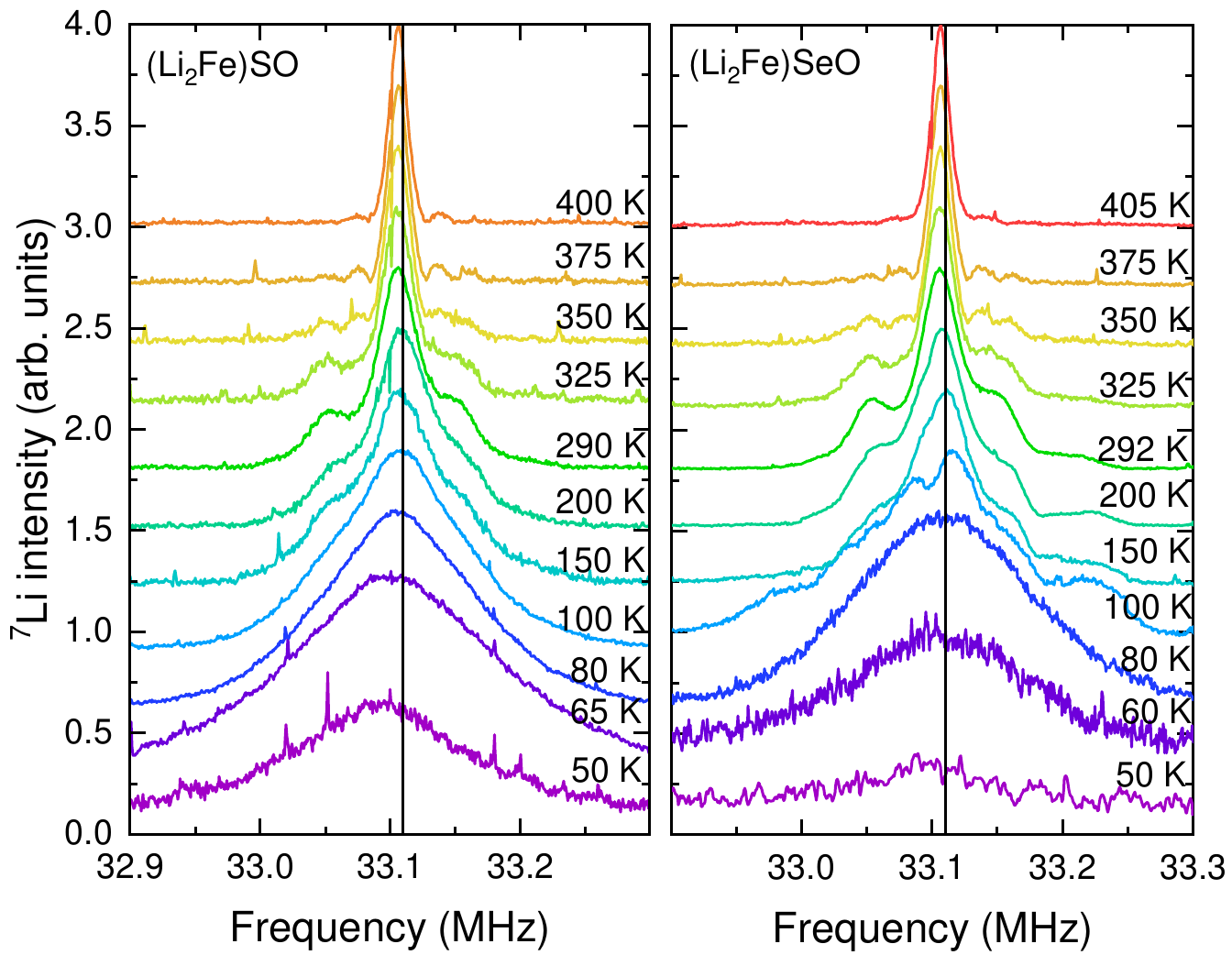}
 \caption{(Color online) The \li\ NMR spectra of \lifeso\ (left) and \lifeseo\ (right) for different temperatures between 50 and 405~K in a static magnetic field of 2.001~$T$. The solid line represents the unshifted resonance frequency of \li\ with $\gamma = 16.5471$~MHz/T at this magnetic  field.}\label{fig:NMRspec}
\end{center}
\end{figure}

 \li\ NMR spectra of polycrystalline \lifeso\ and \lifeseo\ are shown in Fig.~\ref{fig:NMRspec} for temperatures between 50 and 405~K. The spectra exhibit up to 6 separated peaks at high temperatures, which are depicted in detail for both compounds at 375~K and 200~K in Fig.~\ref{fig:NMRspecdetail} in the Appendix. The peaks broaden and overlap when decreasing the temperature. At temperatures below 80~K only a single broad peak is observed which is associated with the onset of static antiferromagnetic correlations and the antiferromagnetic order observed below 50 K. 
 
 The appearance of satellite peaks as evidenced in Fig.~~\ref{fig:NMRspecdetail} can have two reasons: (1) The separated peaks at high temperatures originate from different local magnetic field values due to different numbers and geometries of nearest neighbor {\em paramagnetically polarized} Fe ions in the Li/Fe disordered lattice. This scenario is discussed above for the zero field Moessbauer spectra in the magnetically ordered state below 50~K. (2) An alternative interpretation based on electric quadrupole hyperfine interaction would  produce, in a powder sample for a nuclear spin $I = 3/2$, a central peak accompanied by four equally spaced singularities. The inner two of these singularities would look like horn-shaped, asymmetric peaks, and the two outer singularities would appear as a step-shaped decrease in intensity.~\cite{LangPRB2005} For the materials studied at hand, this is not observed at high temperatures (see Fig.~\ref{fig:NMRspec} and Fig.~\ref{fig:NMRspecdetail}). Even considering different Li sites with different quadrupolar coupling constants would most likely not result in the observed spectral peaks. For explaining at least 5 of the observed 6 line patterns, two different electric field gradients (EFGs) would have to be present with one being approximately twice as large as the other. Such a large difference in the EFG in one compound is indeed difficult to justify. Moreover, the  intensity of the single peaks increases with decreasing temperature and the spectra become asymmetric at low temperatures (see, e.g., the spectra at 292~K and below), whereas any quadrupolar splitting is expected to be symmetric around the central resonance line,  and the intensity of the quadrupolar peaks is not expected to increase with decreasing temperature. Therefore, we conclude that the first scenario is more appropriate. 
 
 The probability distribution for the number of nearest neighbor Fe ions for a stochastic Fe-Li distribution is shown in table~I. This distribution indeed suggests 6 to 7 different Li environments, as e.g. seen in the NMR spectra between 200~K and 375~K. Note that in NMR at this elevated temperatures we study a {\em paramagnetically polarized} Fe spin environment in a powder sample. The internal field distribution is very different from the field distribution in the {\em antiferromagnetically ordered zero field measurements} in Moessbauer spectroscopy. Qualitatively, the high temperature NMR measurements are consistent with the low temperature Moessbauer results.


\begin{figure}[h!]
\begin{center}
 \includegraphics[width=\columnwidth,clip]{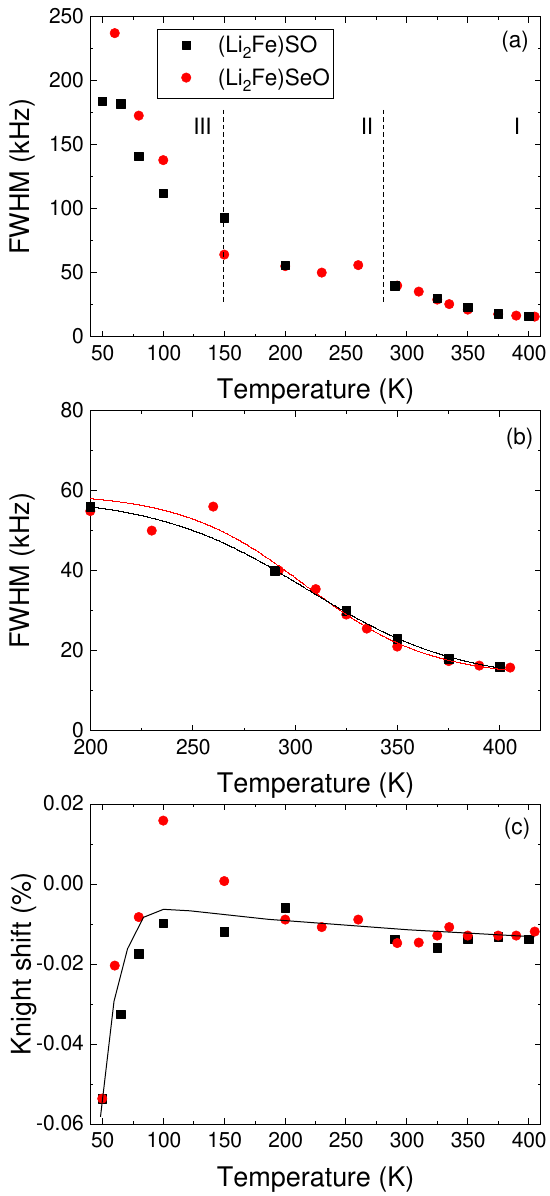}
 \caption{(Color online) (a) FWHM linewidth of the NMR spectra. The dashed lines separate the different temperature ranges mentioned in the text. (b)  High temperature part from (a) with a fit to the temperature dependence of the linewidths to Eqn.~\ref{eq:Tinf}. (c) Knight shift determined from the central peak maximum (open symbols) with a line as a guide to the eye.}
\label{fig:NMRparam}
\end{center}
\end{figure}
 
We determined the Knight shift from the frequency of the central peak maximum of each spectrum and the linewidth as the full width at half maximum (FWHM) from this peak. Due to the presence of multiple peaks, which gain intensity and broaden with decreasing temperature, below 200~K the side peaks contribute to the FWHM of the central peak. 
In the Appendix we present a comparison with the results of an alternative analysis, in which we determined the Knight shift from the center of gravity (COG) of the full spectra, and the linewidth from the square root of the second moment, $\sigma$, of the full spectra (see Fig.~\ref{fig:NMRparam2} and the text). This approach takes into account the contribution of satellite peaks. However, both approaches give qualitatively consistent results. 

The temperature dependence of the linewidth and the Knight shift are shown in Fig.~\ref{fig:NMRparam}. The Knight shift is nearly temperature independent between 405~K and 100~K. Below $\approx 100$~K, $K$ decreases strongly which signals the onset of antiferromagnetic correlations between the Fe electronic spins. 

Regarding the NMR linewidth, we identify three temperature ranges of interest: (I) an increase of the linewidth with decreasing temperature between 400 and 280~K (enlarged in Fig.~\ref{fig:NMRparam} (b)), (II) a constant linewidth between 280 and about 150~K, and (III) a strong increase of the linewidth between 150~K and 50~K. Both compounds, \lifeso\ and \lifeseo\ behave very similar. 

The $T$-dependence of the linewidth in range (I) can most likely be explained by motional narrowing, where the thermally activated hopping of the Li ions averages out the hyperfine fields at the Li nucleus and lead to a narrow linewidth of about 15~kHz at 400~K. Note, that this linewidth is still a factor of $\approx 10$ larger compared to other non-magnetic Li ion battery materials~\cite{Baek2014,Nakamura2006,Kuhn2011}. We attribute this to the magnetic moments of the Fe ions, which generate a large hyperfine field at the Li nucleus even at high temperatures in comparison with non-magnetic Li ion battery materials, where the linewidth is caused  by \li\ homonuclear dipolar broadening of the nuclear spins. 

The Fe magnetic moments do not affect the hopping of Li ions itself but increase the linewidth of the NMR spectra. Since the side peaks also vanish above 400~K, we assume that the hopping rate of all Li ions is  faster than the linewidth, i.e., the inverse correlation time of the Li ion hopping ${\tau}^{-1} \gg 15$~kHz (fast motional regime) above 400~K. 

The onset temperature $T_{onset}$ of the Li ion hopping is the temperature that separates the Li diffusion regime (I) from the rigid lattice linewidth regime (II), in which the linewidth has a constant value of $\approx 60$~kHz.  Below $T_{\textrm{onset}}$, all Li ions are immobile on the timescale of NMR. We estimate $T_{\textrm{onset}} \approx 280$~K. We can describe the temperature dependence of the linewidth in this transition with a phenomenological Fermi-function  Eqn.~\ref{eq:Tinf}. Here, $T_{\textrm{inflection}}$ marks the midpoint and $A$ the width of the transition.   
\begin{equation}
\Delta \nu (T) = \Delta \nu_{\textrm{rl}} - \frac{\Delta \nu_{\infty} - \Delta \nu_{\textrm{rl}}}{1+\exp(\frac{T_{\textrm{inflection}}-T}{A})} \,,
\label{eq:Tinf}
\end{equation}
This analysis results in values of $T_{\textrm{inflection}} = 305$~K, $\Delta \nu_{\textrm{rl}} = 59$~kHz, and $\Delta \nu_{\infty} = 13$~kHz for both compounds. The width of the transition is $A = 37 \pm 4$~K for \lifeso , and $A = 30 \pm 8$~K for \lifeseo . These parameters agree with the above determined values, e.g. $T_{\textrm{onset}} \approx T_{\textrm{inflection}} - A/2$, and the linewidth determined from the spectra at 400~K corresponds to $\Delta \nu_{\infty}$.

According to Waugh and Fedin~\cite{Waugh1963,Baek2014} one can correlate the onset temperature $T_{onset}$ to the activation energy $E_a$ of the Li ion diffusion process by
\begin{equation}
E_a~\mathrm{(eV)} \sim 1.67 \cdot 10^{-3}~T_{onset}~\mathrm{(K)} \,,
\label{eq:Ea}
\end{equation}
which gives an approximate $E_a$ value of 0.47~eV. This value is slightly higher than the value from DFT calculations (0.32~eV) \cite{Lu.2018} and comparable to other Li ion battery materials~\cite{Nakamura2000,Verhoeven2001,Pecher2017,Neef2020,Kuhn2011,Baek2014}.

Further cooling below $\approx 150$~K leads to an additional broadening of the \li\ spectra in range (III) which we attribute to slowing down of Fe spin fluctuations and subsequently appearance of static magnetic hyperfine fields at the Li sites, similar to \li\ NMR in LiNiO$_2$~\cite{Nakamura2000}. This marks the onset of short range static magnetic order, which cannot be distinguished from long range magnetic order setting below 50~K. Due to the large linewidth the NMR spectra in the ordered state do not reveal detailed information  in the Li/Fe site disordered compounds \lifeso\ and \lifeseo.

\subsubsection{Spin-spin relaxation}
\label{sec:NMRT2}

\begin{figure}[htb]
\begin{center}
 \includegraphics[width=\columnwidth,clip]{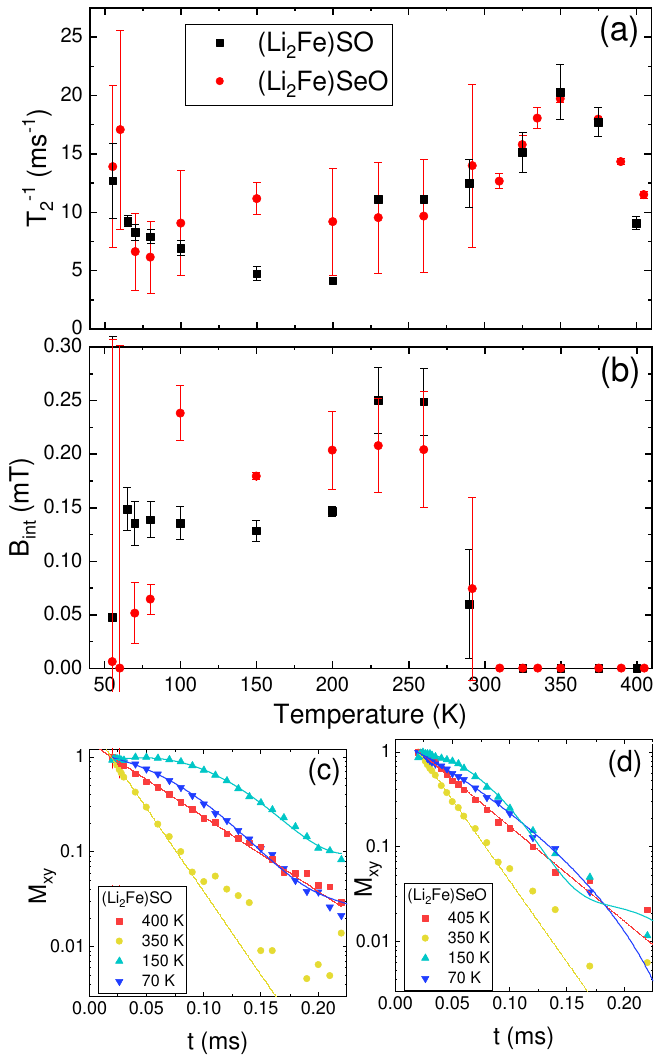}
 \caption{(Color online) (a) \ssrr\ for \lifeso\ (black squares) and for \lifeseo\ (red circles) versus temperature. (b) Internal field $B_{\textrm{int}}$ determined from Eqn.~\ref{eq:T2}. (c) Spin-echo decay of $M_0$ versus $\tau$ for \lifeso , and (d) for \lifeseo .}
\label{fig:NMRT2}
\end{center}
\end{figure}

In Li ion battery materials, in the slow motion regime, where $\omega_L\tau_c > 1$ with $\omega_L$ and $\tau_c$ being the nuclear Larmor frequency and the Li diffusion rate, respectively, 
the spin-spin relaxation rate \ssrr\ typically follows the temperature dependence of the linewidth~\cite{heitjans2006,Baek2014}. In this temperature range the NMR linewidth reflects the static distribution of local fields, and the nuclear spins start to de-phase in these local fields over time. Therefore, as temperature decreases and motion slows down, linewidth and \ssrr\ increase in a correlated way. However, as can be seen in Fig.~\ref{fig:NMRT2}, for the materials studied in this work this holds only for temperatures above 350~K. 

Below 350~K, \ssrr\ decreases with decreasing temperature, whereas the linewidth continues to increase (see Fig.~\ref{fig:NMRparam}). Furthermore, indications for an oscillatory behavior set in in the recovery curves $M_{xy}(\tau)$ vs. $\tau$ below about 300~K (see Figs.~\ref{fig:NMRT2}(c) and (d)). 
Such oscillations in nuclear spin-spin relaxation can be caused
by interaction with a very small static internal or external magnetic field. We therefore fit the relaxation of the nuclear magnetization to the following equation~\cite{Lombardi1996,Straessle2011,Boughama2022}:

\begin{equation}
M_{xy}(\tau) = M_0 \cdot\exp(-(2\tau)/T_2) \cdot (1+F \cdot \cos(2\omega_{\textrm{int}} \tau-\phi)) \, ,
\label{eq:T2}
\end{equation}

where $M_0$ is the initial nuclear magnetization, $\tau$ the separation time between the 90\degree\ and the 180\degree\ pulses, $T_2$ the nuclear spin-spin relaxation time, $\omega_{\textrm{int}} = {^7\gamma} \cdot B_{\textrm{int}}$ the frequency of oscillations, and $\phi$ its phase shift. The pre-factor, $F$, determines the relative strength of the oscillatory component in the transverse magnetization $M_{xy}(\tau)$. F is reduced, if the internal field, $B_{\textrm{int}}$, is distributed, i.e. all nuclei precess at slightly different frequencies. In fact, the recovery curves in  Figs.~\ref{fig:NMRT2}(c) and (d) do not show a clean oscillatory behavior with several cycles, but only one low frequency cycle. This is primarily due to the overall short $T_2$, but also due to the reduction of $F$ in a powder sample that is about $1/3$ on the whole temperature range, where the oscillations can be observed. Note that other functions such as a Gaussian relaxation do not fit the data.

From the fits, we extract the internal magnetic exchange field, $B_{\textrm{int}}$ that is shown in Fig.~\ref{fig:NMRT2}(b). It amounts to $B_{\textrm{int}} \approx 0.15$~mT, which corresponds to a Larmor frequency of $\omega_{\textrm{int}} \approx 2.5$~kHz. This is well below the linewidth of the central transition at all temperatures. Therefore, this internal magnetic exchange field does not lead to line broadening, but only to oscillations in the nuclear spin-spin relaxation. 

The small internal field at the Li ions may arise from 
small 
magnetic Fe impurity clusters, as detected also in the macroscopic susceptibility and Moessbauer measurements (see above). This would agree to the fact that the oscillations are more clearly visible in the S sample than in the Se sample (compare Fig.~\ref{fig:NMRT2}(c) and (d), and the larger error bars in the fit results for the Se sample), because the impurity content of the S samples seems to be larger than for Se. Despite the fact that the fits are not very accurate and the error bars are relatively large, the observed internal magnetic field of only 0.15~mT could well be explained by a minor phase of ferromagnetic impurities.

The appearance of a small static internal magnetic field could also explain the decrease of \ssrr\ below 350~K. In the motional narrowing regime, \ssrr\ is dominated by fast fluctuations of the local fields, i.e., by fast motion of the Li ions. If the fluctuations slow further below 350 K, they might enter a regime where they no longer efficiently contribute to transverse relaxation, causing \ssrr\ to decrease while the linewidth still broadens due to a broader distribution of static or quasi-static local fields. This could indicate a crossover from a fast fluctuating regime to a more inhomogeneous static-like regime. The emergence of oscillatory behavior in $M_{xy}(t)$ suggests the onset of local static fields from the ferromagnetic impurity, causing the $T_2$ relaxation transitioning from fluctuation-driven to coherence-driven, where nuclear spins precess in quasi-static internal fields rather than relaxing due to fluctuating ones. This could reduce \ssrr\ despite the further increase in linewidth. It also shows that the measurement of Li ion diffusion by NMR in battery materials containing magnetic ions can be difficult and does not always lead to unambiguous results.

\subsubsection{Spin-lattice relaxation}
\label{sec:NMRT1}

The spin-lattice relaxation rate $1/T_1$ is determined by the magnetic hyperfine field amplitude oscillating with the nuclear Larmor frequency $\omega_L$. In \lifeso\ and \lifeseo\  this amplitude is due to electronic spins of the nearest neighbor Fe sites. $1/T_1$  measures the imaginary part of the dynamical spin susceptibility $\chi^{\prime\prime}$ of the electronic spin system at $\omega_L$: 
\begin{equation}
(T_1T)^{-1} \propto \sum_{\vec{q},\alpha,\beta}F_{\alpha,\beta}(\vec{q}) \frac{\chi^{\prime\prime}_{\alpha,\beta}(\vec{q},\omega)}{\omega} \, .
\label{eq:T1T}
\end{equation}
Here, $F_{\alpha,\beta}(\vec{q})$ denotes the hyperfine form factors with $\alpha,\beta = {x,y}$ that are determined by the Fourier transformation of the hyperfine coupling tensor. If a system approaches a magnetic order, the dynamical spin susceptibility is enhanced, which leads to an enhancement of \slrr\ or \slrrt , respectively. Below the magnetic order transition, the spin fluctuations decrease again, and so does \slrr , leading to a maximum in \slrr\ at the magnetic ordering temperature.

For \slrr\ measured by NMR on powder samples, if the quadrupole splitting is not large and all transitions fall into one resonance line, the recovery function is given by $M_z(t) = M_0[1-f(\exp(-(t/T_1)^\beta)$, where $f=1$ for an ideal saturation of the nuclear magnetization, $M_0$. $\beta$ is a stretching parameter that describes a distribution of spin lattice relaxation times, which often occurs in disordered systems, where nuclei located in different environments obey different relaxation times. 

\begin{figure}
\begin{center}
 \includegraphics[width=\columnwidth,clip]{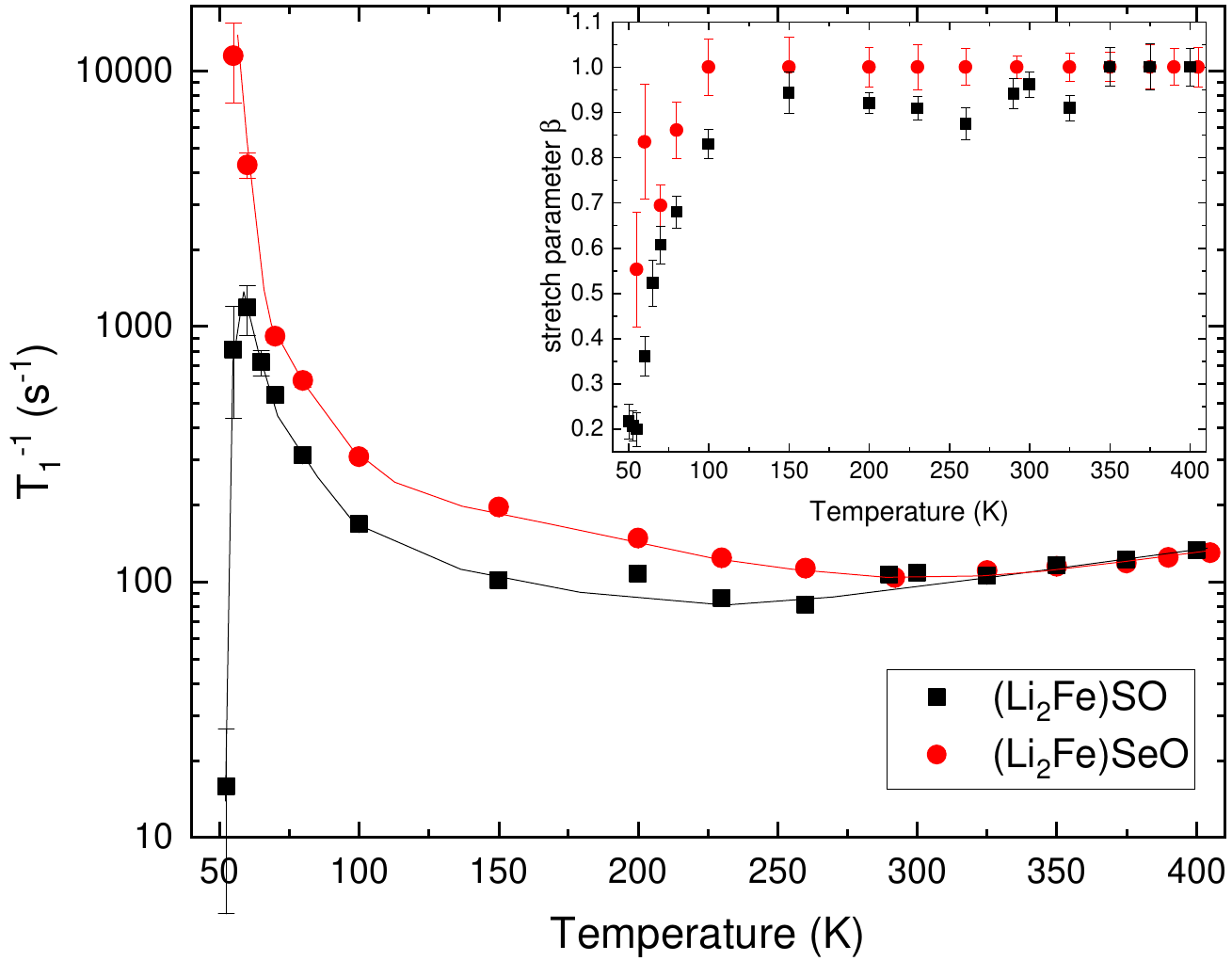}
 \caption{(Color online) \slrr\ for \lifeso\ (black squares) and for \lifeseo\ (red circles) versus temperature. The inset shows the exponent (stretching parameter) $\beta$. The lines are guides to the eye.}
\label{fig:NMRT1}
\end{center}
\end{figure}

\slrr\ versus temperature is shown in Fig.~\ref{fig:NMRT1} for \lifeso\ and for \lifeseo . Similar to the linewidth data, we can identify three interesting temperature ranges: (i) an increase of \slrr\ between 290 and 400~K. This is most likely related to the onset of Li ion diffusion, which induces additional relaxation through random fluctuation of local magnetic fields at the Li nucleus. In this temperature range, the relaxation can be described by the Bloembergen, Purcell, Pound (BPP) model~\cite{Bloembergen1948,Buschmann2011,heitjans2006} 
\begin{equation}
(T_1)^{-1} \propto \frac{\tau_c}{1+\omega_L^2 \tau_c^2} \, ,
\label{eq:BPP}
\end{equation}
where $\tau_c = \tau_0 \exp(E_a/k_B T)$ is the correlation time of the Li ion motion which is coupled to the Li nuclear spin-lattice relaxation by dipolar interaction. $E_a$ is the activation energy, which is the same as the activation energy determined above from the temperature dependence of the linewidth. In this model, the nuclear relaxation due to Li ion motion exhibits a maximum if the condition $\omega_L \tau_c=1$ is met. Since \slrr\ does not exhibit a maximum below 400~K, we conclude that the correlation time of the Li motion is larger than the inverse Larmor frequency  in the measured temperature range, i.e. $\omega_L \tau_c \gg 1$. 

We can use the activation energy $E_a=0.47$~eV determined from the temperature dependence of the linewidth, and estimate the correlation time $\tau_c$ at the inflection point $T_\textrm{inflection}$ from the rigid lattice linewidth, $\tau_c(T_\textrm{inflection}) = 1/\Delta \nu_{\textrm{rl}} = 17 \cdot 10^{-6}$~s. This gives $\tau_0 = \tau_c(T_\textrm{inflection})/\exp(E_a/k_B T_\textrm{inflection}) = 3.14 \cdot 10^{-13}$~1/s. With this data, we estimate a maximum in \slrr\ at a temperature of $T_\textrm{max} = E_a/k_B \ln(\tau_c(T_\textrm{max})/\tau_0 = 475$~K, which is clearly above the measured temperature range up to 405~K. While our limited temperature range does not allow any conclusions about the type of Li ion diffusion (continuum diffusion, jump diffusion, etc.) due to the missing maximum, it does explain the increase of \slrr\ above 290~K, which can be related to the diffusion of the Li ions.    

Below 100~K the spin lattice relaxation rate starts to increase. For \lifeso , \slrr\ exhibits a maximum at $\approx 60$~K, which we identify as the onset of magnetic ordering, similar to the results of macroscopic susceptibility and Moessbauer spectroscopy. In contrast, \slrr\ of \lifeseo\ continues to increase until the \li\ signal disappears. This loss of signal intensity is visible in the NMR spectra in Fig.~\ref{fig:NMRspec}, where the noise level  at 80~K is already much stronger for \lifeseo\ compared to the spectra of \lifeso . At lower temperatures in the ordered state, when the spin lattice relaxation rate is expected to slow down, static hyperfine field distributions in the Li/Fe site disordered systems could lead to such a broadening that one cannot obtain the spectra anymore. The disorder further contributes to the linewidth so that it is more and more difficult to obtain a reasonable signal. For \lifeseo , a phenomenological Curie-Weiss-fit \slrrt $= C/(T-\Theta)$ gives an ordering temperature of $\Theta = 54$~K. This value of $\Theta$ agrees with the temperature where the minimum in $\partial (\chi T)/\partial T$ is observed (see Fig.~\ref{cpmag}).

Another characteristics  of the relaxation curves is a stretched exponential relaxation, which indicates a distribution of relaxation rates that can be described by a stretching exponent $\beta$ in the relaxation function~\cite{johnston2006}. $\beta$ is shown in the inset of Fig.~\ref{fig:NMRT1}. It is about 1 down to 100~K, which means no distribution of relaxation and all nuclei relax with the same $T_1$. Below 100~K, where short range magnetic order sets in and leads to fluctuating hyperfine fields, $\beta$ strongly decreases, indicating that different Li nuclei experience different levels of fluctuating hyperfine fields, which in turn agrees with the site disorder of Li and Fe atoms.

\section{Final discussion and summary}

We report magnetic susceptibility, magnetization, Mössbauer spectroscopy, and NMR studies on \lifeseo\ and \lifeso . Our data reveal a Pauli paramagnetic-like behavior below 300\,K, a long-range antiferromagnetically ordered ground state below $\approx$ 50~K and a regime of short-range magnetic correlations up to 100~K. This reflects the strong dilution of the paramagnetic iron ions on the TM site as well as the geometric frustration of the antiferromagnetic nearest neighbor magnetic exchange interactions on the fcc TM sublattice with dominant triangular magnetic exchange geometry.
Our results are consistent with a random Li-Fe distribution on the shared lattice position so that percolation effects must be taken into account. The 1/3 population of the TM site by magnetic iron only slightly exceeds the theoretical percolation threshold on a cubic lattice of $\simeq 0.31$~\cite{Bocquet1994}. 
On the other hand, introducing nonmagnetic sites into the {individually} frustrated and highly degenerated Kagome planes provides the random combination of frustrated and unfrustrated plaquettes which can favor long-range antiferromagnetic order as observed at hand, spin-glass behaviour or intriguing order-disorder phenomena~\cite{Villain1979,Marinari2000}. The geometric frustration 
manifests itself also in a non-linear saturation effect 
in the dependence of the $^{57}$Fe magnetic hyperfine field on the number $n$ of nearest neighbor Fe ions in the long-range ordered state.  

Depending on the local geometry of the occupied and unoccupied paramagnetic sites, magnetic frustration can be generated and lifted. This may yield the tendency to stabilize long-range magnetic order and to increase the ordering temperature. 
On the other hand, increasing dilution will weaken and eventually completely suppress long-range magnetic order when exceeding the percolation threshold for the given systems which has been shown for several related examples~\cite{Beath2006,Jiang1992,Huecker1999,Yasuda2001}. Our observation of two regimes of short- and long-range magnetic order in \lifeseo\ and \lifeso\ reflect these contrary tendencies. 

Thermally activated Li-ion hopping is deduced by motional narrowing of the \li\ NMR resonance line above approximately 250~K. From the temperature dependence of the linewidth, we estimate an activation energy of $E_{\rm a} = 0.47$~eV. The onset of Li-ion motion also leads to an increase in the spin-lattice relaxation rate \slrr\ with increasing temperature. However, a maximum in \slrr\ could not be observed within the accessible temperature range of our experiment. The estimated activation energy is similar to values reported for other Li-ion battery materials, making lithium-rich antiperovskites a promising cathode material.

In general, our study elucidates magnetism in a disordered semimetallic system with geometric frustration and thermally induced ion diffusion dynamics.


\begin{acknowledgements}
The authors gratefully acknowledge valuable discussion with Samuel Coles, Jochen Geck, Martin Valldor, and Maurits Haverkort.
Work has been done in the framework of the joint project KL 1824/20-1 \& GR 5987/2-1 funded by Deutsche Forschungsgemeinschaft (DFG). Technical support and discussions within the DFG Research Training Group “Mixed Ionic Electronic Transport” (GRK 2948) are gratefully acknowledged. F.~S. and H.-H. K acknowledge funding by the DFG collaborative research center SFB 1143 "Correlated Magnetism: From Frustration to Topology". Work has also been supported within the framework of the Excellence Strategy of the Federal and State Governments of Germany via the Heidelberg University's flagship EMS initiative and the Cluster of Excellence STRUCTURES, and via the TU Dresden/U Würzburg Cluster of Excellence ct.qmat. M.A.A.~M. thanks the IFW excellence program for financial support. 
\end{acknowledgements}

\bibliography{litLi2FeXO}

\hfill
\newpage

\section{Appendix}
\label{sec:appendix}

\renewcommand{\thefigure}{S\arabic{figure}}
\setcounter{figure}{0}

\subsection{Detail of the NMR spectra}

\begin{figure}[h]
\begin{center}
 \includegraphics[width=0.95\columnwidth,clip]{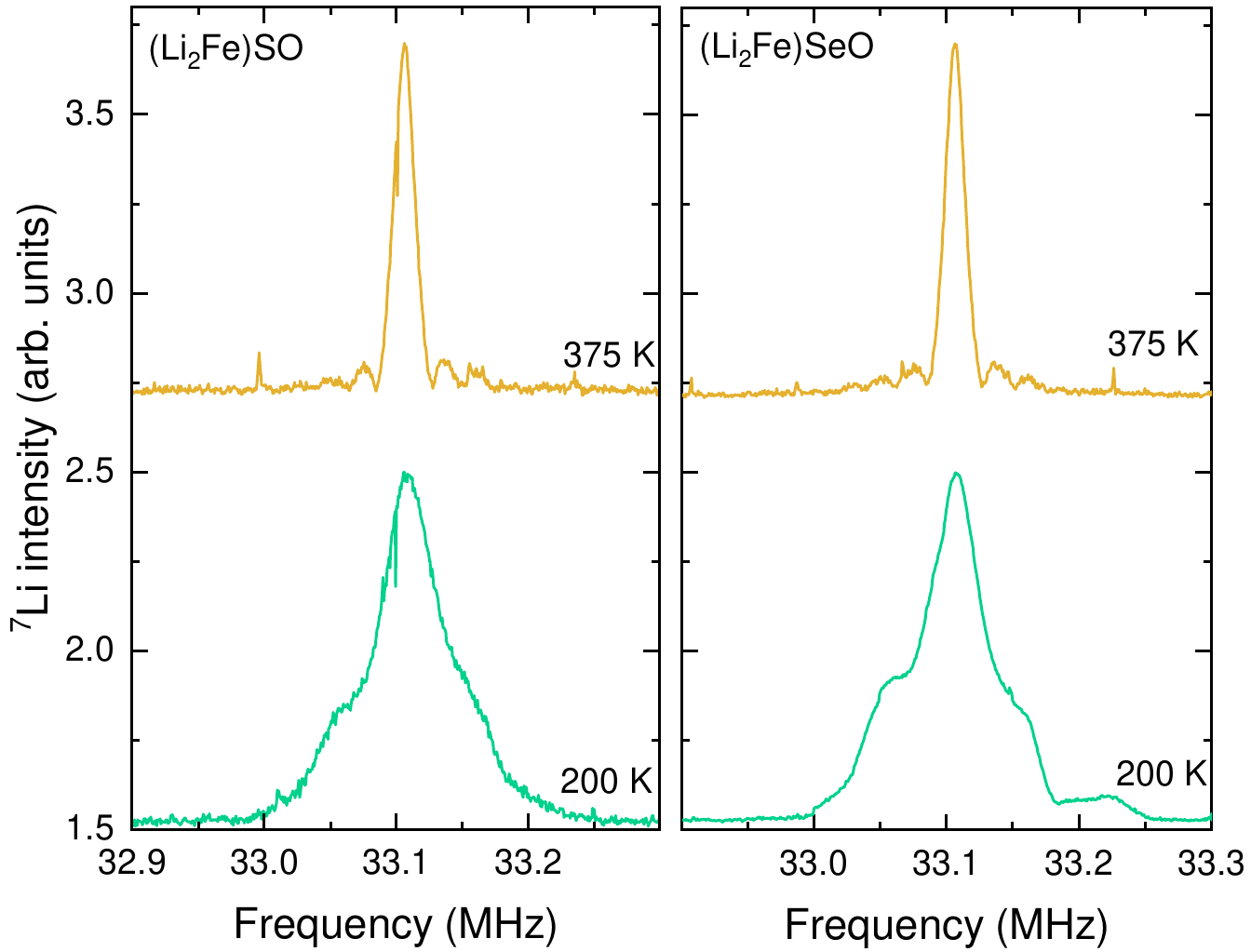}
 \caption{(Color online) The $^7$Li NMR spectra of \lifeso\ and \lifeseo\ at 375~K and at 200~K.}
\label{fig:NMRspecdetail}
\end{center}
\end{figure}

Figure~\ref{fig:NMRspecdetail} shows the $^7$Li NMR spectra of \lifeso\ and \lifeseo\ at 375~K and at 200~K.  
At high temperatures, the spectra show up to six separated peaks. The intensity of these additional peaks increases with decreasing temperature, and the all peaks broaden at lower temperature. Moreover, the spectra become asymmetric at low temperatures.

\subsection{Comparison of two strategies to determine NMR linewidth and Knight shift}

It is not straightforward to determine the linewidth ($\Delta \nu$) and the Knight shift ($K$) consistently for all temperatures. In the main text, We determined the Knight shift from the frequency of the central peak maximum of each spectrum and the linewidth as the full width at half maximum (FWHM) from this peak. Due to the presence of multiple peaks, which gain intensity and broaden with decreasing temperature, below 200~K the side peaks contribute to the FWHM of the central peak. On the other hand, if we read the linewidth from the central peak, we do not take into account the contribution of the side peaks to the spectra at high temperatures. Here, we present a comparison with the results of an alternative analysis, in which we determined the center of gravity (COG) of the full spectra and the square root of the second moment, $\sigma$, of the full spectra. Both approaches give qualitatively consistent results. 

\begin{figure}[h!]
\begin{center}
 \includegraphics[width=0.95\columnwidth,clip]{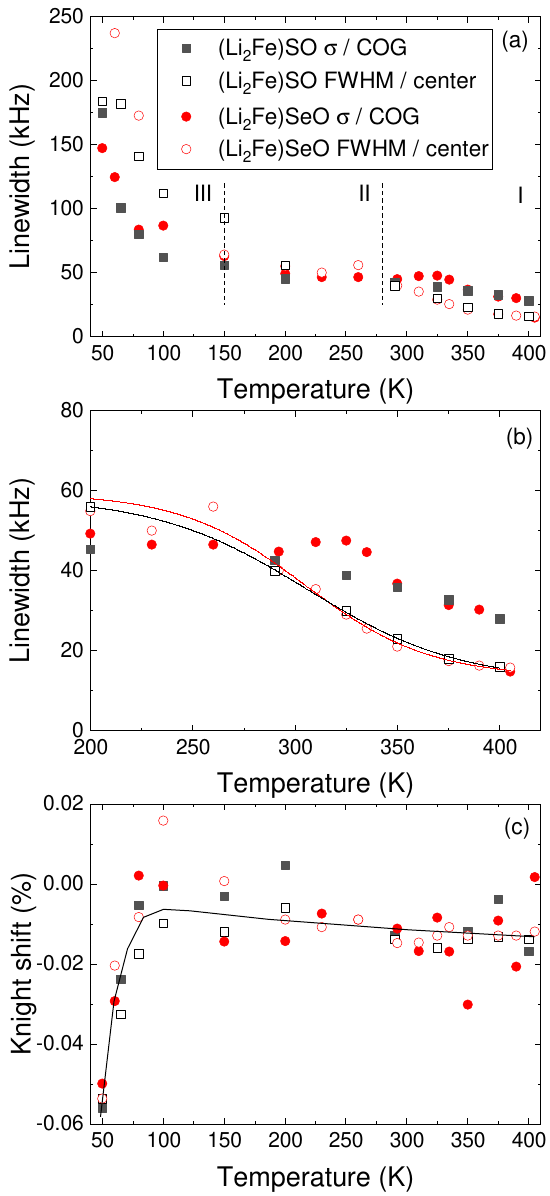}
 \caption{(Color online) (a) The linewidth of the NMR spectra. Closed symbols are determined from the center of gravity (COG) and the square root of the second moment, $\sigma$. Open symbols are read from the maximal peak (center frequency) and the full width at half maximum (FWHM). Black symbols are for \lifeso , red ones for \lifeseo . The dashed lines separate the different temperature ranges mentioned in the text. (b) Detail of the high temperature part from (a) with a fit to the temperature dependence of the linewidths (FWHM of center peaks) to Eqn.~\ref{eq:Tinf}. (c) Knight shift from COG (closed symbols) and maximal peak (open symbols) with a line as a guide to the eye.}
\label{fig:NMRparam2}
\end{center}
\end{figure}

\end{document}